\documentclass{ws-ijwmip}
\usepackage{graphicx}
\usepackage[super]{cite}
\usepackage{cite}
\usepackage{amsmath,amssymb,amsfonts}
\usepackage{graphicx}
\usepackage{textcomp}
\usepackage{xcolor}
\usepackage{import}
\usepackage{algpseudocode}
\usepackage{algorithm}
\usepackage{algorithmicx}
\usepackage{subcaption}
\usepackage{multicol}
\usepackage{booktabs}
\usepackage{multirow,fixltx2e}
\usepackage{multicol}
\usepackage{hhline} % For \cline use \hhline{~-----}
\usepackage{float}
\usepackage{url} % for url in biblography
\usepackage{textcomp} % For 's
\usepackage{hyperref}
\usepackage{balance}
\usepackage{dcolumn}
\begin{document}

\markboth{Rohit Agrawal}{$\ell_1$SABMIS}

%%%%%%%%%%%%%%%%%%% Publisher's Area please ignore %%%%%%%%%%%%%%%%%%%%%%%
\catchline{}{}{}{}{}
%%%%%%%%%%%%%%%%%%%%%%%%%%%%%%%%%%%%%%%%%%%%%%%%%%%%%%%%%%%%%%%%%%%%%%%%%%

\title{$\ell_1$SABMIS: $\ell_1$-minimization and sparse approximation based blind multi-image steganography scheme}

\author{Rohit Agrawal}
\address{Data \& Computational Sciences Laboratory, Indian Institute of Technology Indore, \\
Indore, 453552, India\\
phd1501201004@iiti.ac.in}

%\author{Kapil Ahuja}

%\address{Data \& Computational Sciences Laboratory, Indian Institute of Technology Indore, \\
%Indore, 453552, India\\
%kahuja@iiti.ac.in}

\maketitle

%\begin{history}
%\received{(Day Month Year)}
%\revised{(Day Month Year)}
%\accepted{(Day Month Year)}
%\published{(Day Month Year)}
%\comby{(xxxxxxxxxx)}
%\end{history}

\begin{abstract}
Steganography plays a vital role in achieving secret data security by embedding it into cover media. The cover media and the secret data can be text or multimedia, such as images, videos, etc. In this paper, we propose a novel $\ell_1$-minimization and sparse approximation based blind multi-image steganography scheme, termed $\ell_1$SABMIS. By using $\ell_1$SABMIS, multiple secret images can be hidden in a single cover image. In $\ell_1$SABMIS, we sampled cover image into four sub-images, sparsify each sub-image block-wise, and then obtain linear measurements. Next, we obtain DCT (Discrete Cosine Transform) coefficients of the secret images and then embed them into the cover image\textquotesingle s linear measurements.

We perform experiments on several standard gray-scale images, and evaluate embedding capacity, PSNR (peak signal-to-noise ratio) value,  mean SSIM (structural similarity) index, NCC (normalized cross-correlation) coefficient, NAE (normalized absolute error), and entropy. The value of these assessment metrics indicates that $\ell_1$SABMIS outperforms similar existing steganography schemes. That is, we successfully hide more than two secret images in a single cover image without degrading the cover image significantly. Also, the extracted secret images preserve good visual quality, and $\ell_1$SABMIS is resistant to steganographic attack.
\end{abstract}


%\ccode{AMS Subject Classification: 22E46, 53C35, 57S20}

\section{Introduction}\label{introduction}
The security of digital data is essential for its transfer over the communication media. To achieve this data security, in general, there are mainly two approaches used; cryptography \cite{Stalling} and steganography \cite{Steg_Rohit}. In cryptography, the encryption mechanism transforms plain-text (i.e., secret data) into cipher-text using the encryption key. This cipher-text appeared as an unreadable form that attracts opponents to manipulate its contents by using certain brute-force attacks \cite{Stalling}. Nevertheless, steganography avoids this situation. 

Steganography is derived from two Greek words; steganos means ``covered or secret," and graphie means ``writing". The purpose of steganography is to hide the secret data into some other unsuspected cover media so that the secret data becomes visually imperceptible. In steganography, both the cover media and the secret data can be text or multimedia. The media obtained after embedding secret data into cover media is called stego-media. In this paper, we consider both the secret data and the cover media as images due to their heavy use in web-based applications. The challenges here are: enhancing the embedding capacity, preserving the quality of the stego-image, and the scheme should be resistant to steganalysis\footnote{It is the study of detecting the secret data hidden using steganography.} (i.e., steganographic attacks). In the following paragraphs, first, we discuss different categories of steganography schemes. Second, we discuss some of the existing schemes and their weaknesses. Finally, we discuss how the scheme proposed in this paper outperforms the existing schemes. 

In general, image steganography can be categorized based on the domain in which embedding is performed, i.e., the spatial domain or the transformation domain. In the spatial domain-based scheme, secret data is embedded directly into the cover image by some alteration in image pixels value \cite{Abbas, Chan1, Wang, Chan2, Steg_2019_FGT}. In the transform domain-based scheme, initially, the cover image is transformed into frequency components by certain transformations, and then the secret data is embedded into these components. A few of those schemes are JSteg \cite{Jsteg}, F5 \cite{F5}, Outguess \cite{Outguess}, etc \cite{Chang, Liu, Pan, Steg_World_Scienticfic, Steg_Rohit}. The steganography scheme can also be categorized based on their embedding mechanisms, i.e., direct embedding or indirect embedding. In a direct embedding mechanism, the secret data is embedded by flipping the LSB (Least Significant Bit) of either the pixel value (i.e., spatial domain-based schemes) or the transformed coefficients (i.e., transform domain-based schemes) of the cover image \cite{Steg_2019_FGT, Jsteg, Outguess, Chang}. In an indirect embedding mechanism, the pixel value or the transformed coefficient values are altered according to certain secret message bits \cite{F5, Liu, Pan, Steg_Rohit}. 

The spatial domain-based image steganography schemes outperform the transform domain one in-terms of embedding capacity. But they are not resistant to steganographic attack. Transform-based schemes are resistant to these attacks and provide good visual quality stego-image with limited embedding capacity. Furthermore, if we try to increase the embedding capacity of these schemes, then the quality of stego-image degrades. Besides, the indirect embedding mechanisms make steganography schemes more resistant to these attacks \cite{Westfeld, Fridrich, Zhang, Nissar, Patsakis} as compared to direct ones. Thus, some of the schemes mentioned above are not resistant to steganographic attack.\cite{Jsteg, Outguess, Chang} While some are resistant to these attacks. \cite{F5, Liu, Pan, Steg_Rohit} 

%%%% Literature %%%%%
The purpose of the scheme proposed in this paper is to hide multiple secret images in a single cover image. Hence, first, we discuss some recent image steganography schemes that hide multiple secret images in a single cover image. In \cite{S_Hemalatha}, the authors proposed a steganography scheme in which they hide two gray-scale secret images in a single color image. This scheme is based on two-level DWT (Discrete Wavelet Transformation). In this, the quality of stego-image and extracted images are good, but the embedding capacity in bit per pixel is very less. In \cite{Prashanti}, the authors proposed a steganography scheme in which they hide three binary secret images in a single gray-scale as well as in a color image. This scheme is based on LSB based embedding approach, which is not resistant to steganographic attack. In this, the quality of stego-image is good, but the embedding capacity in bit per pixel is very less. 

Now, we discuss some recent image steganography schemes that hide only one secret image in a single cover image. In \cite{Sanjutha_MK}, a gray-scale secret image is embedded in a color image. This scheme is based on DWT and PSO (particle swarm optimization). In \cite{S_Arunkumar_MedicalImage_JIFS}, a binary medical image is embedded in to a gray-scale cover image. This scheme is based on Redundancy Integer Wavelet Transform (RIWT), SVD (Singular Value Decomposition), and Discrete Cosine Transformation (DCT). In \cite{S_Arunkumar_MedicalImage}, a binary medical image is embedded in a color cover image. This scheme is based on RIWT, DCT, and QR factorization. In all these schemes, the stego-image preserved good visual quality, but their embedding capacity is limited. And, if we try to increase this embedding capacity, the quality of the stego-image degrades. 

To overcome this limitation, in this manuscript, we utilize other paradigms; optimization (i.e., $\ell_{1}$-minimization) and sparse approximation. The steganography scheme proposed in this manuscript is termed as $\ell_1$SABMIS because we use the concept of $\ell_1$-minimization and sparse-approximation for blind multi-image steganography scheme. $\ell_1$SABMIS fulfills all the requirements of image steganography, i.e., it has high embedding capacity, preserve good visual quality stego-image, and resistant to steganographic attacks. Here, first, the cover image is sampled into four sub-images. Next, each sub-image is sparsified using DCT, and then its linear measurements are obtained using a random measurement matrix. Next, we collect the DCT coefficients of all secret images. Now, we select fix number of these coefficients for each secret image and then embed them into the permissible linear measurements using our proposed embedding rule (discussed in Section \ref{propose method}) and obtain modified measurements. Finally, we generate the stego-image from these modified measurements by solving the $\ell_1$-minimization problem (again, discussed in Section \ref{propose method}). 

In $\ell_1$SABMIS, as discussed earlier, we embed DCT coefficients of secret images into linear measurements of the cover image, instead of embedding them directly into the transformed coefficients. These measurements act as encoded transformed coefficients, and hence, also adds security to our proposed scheme. For performance evaluation, we perform experiments on standard test images and evaluate embedding capacity, PSNR (Peak Signal-to-Noise Ratio) value \cite{Steg_Rohit}, mean SSIM (Structural Similarity) index \cite{SSIM}, NCC (Normalized Cross-Correlation) coefficient \cite{NC}, NAE (Normalized Absolute Error, also called normalized average absolute difference) \cite{NC}, and entropy \cite{Gonzalez}. We also show the visual comparison between the cover image \& its corresponding stego-image, and between the secret image \& its corresponding extracted secret image. Moreover, we also show that our scheme outperforms existing image steganography schemes.

The rest of the paper is organized as follows. Section \ref{Background} gives brief explanation of each of $\ell_{1}$-minimization and sparse approximations. Section \ref{propose method} explains our proposed steganography scheme, including embedding and extraction of the secret image. Section \ref{experimental results} presents experimental results. Finally, Section \ref{conclusion} gives conclusions and future work.

\section{Background}\label{Background}
Here, in the following subsections, we provide a brief explanation of $\ell_1$-minimization and sparse approximation. 
\subsection{$\ell_1$-Minimization}\label{l1_minimization}
The general form of the $\ell_1$-minimization is given as\cite{L1Minimization_World_Scienticfic, Rohit_ParaLarPD} 
\begin{equation}
\begin{aligned}\label{eq: CS reconstruction l1}
& \min_x \left \| x \right \|_{1} \\ %\notag
& \text{Subject to} \hspace{0.1cm} A x = b,
\end{aligned}
\end{equation} 
where $\left \| \cdot \right \|_{1}$ is the $\ell_{1}$-norm, $x\in R^{N\times 1}$, $A\in R^{M\times N}$, and $b\in R^{M\times 1}$. In many applications such as signal processing, pattern recognition, etc., a~$\ell_{1}$-minimization or a~minimum $\ell_{1}$-norm solution is preferable. A few examples of these applications include data separation, face recognition, data clustering, image restoration, image classification, etc.

\subsection{Sparse Approximation}\label{sparse_approximation}
Many practical problems in digital image processing and other disciplines include finding the best approximate solution to a system of linear equations. The theory of sparse approximation corresponds with sparse solutions for linear equation systems. This theory has applications in many domains, such as signal/ image processing, machine learning, medical imaging, etc. 
%\footnote{This signal can be view as a column vector of length \textit{N}}

Let, we have an unknown \textit{K}-sparse signal $x\in R^{N\times 1}$, a measurement matrix $\Phi \in R^{M\times N}$, and a measurement vector $y\in R^{M\times 1}$, such that $y=\Phi x$. Reconstruction (or approximate reconstruction) of $x$ from $\Phi$ and $y$, which is usually considered as an inverse problem, can be obtained by solving the following $\ell_{0}$-minimization problem\cite{SparseApproximation_World_Scienticfic}:
%%%%%%%%%%%%%%%%%%%%%%% L0 minimization equation %%%%%%%%%%%%%%
\begin{equation}
\begin{aligned}\label{eq: CS reconstruction l0}
&\min_x \left \| x \right \|_{0} \\ %\notag 
&\text{Subject to} \hspace{0.1cm} \Phi x = y,
\end{aligned}
\end{equation}
where $\left \| \cdot \right \|_{0}$ is $\ell_{0}$-norm, which measures the total number of non-zero elements in a vector. This equation (i.e., \eqref{eq: CS reconstruction l0}) is referred to as $\ell_{0}$-norm minimization problem,\cite{SparseApproximation_World_Scienticfic} which is a combinatorial and NP-hard problem \cite{Baraniuk}. Since the solution of $x$ is sufficiently sparse, we can substitute the $\ell_{0}$-norm minimization by the $\ell_{1}$-norm (i.e., the closest convex norm) minimization problem. Hence, $x$ is reconstructed from $\Phi$ and $y$ by the solving \eqref{eq: CS reconstruction l1}, where $A$ and $b$ are equivalent to $\Phi$ and $y$, respectively. 

As discussed, the signal to be approximated (or reconstructed) should be sparse. However, there are cases when this signal is not sparse. So, we can sparsify it using some transformation. For example, the signal $x$, which is not sparse, can be \textit{sparsified} using an orthogonal matrix (termed as sparsification matrix) $\Psi\in R^{N\times N}$ as %$x$ is \textit{sparsified} as
\begin{equation}\label{eq:CS sparsification}
	s=\Psi^{T} x,
\end{equation}
where $s\in R^{N\times 1}$ is the sparse representation of $x$. Sparse signal $s$ can be reconstructed by solving \eqref{eq: CS reconstruction l1}, where the decision variable $x$ is equivalent to $s$, $A$ is equivalent to $\Theta = \Phi \Psi$, and $b$ is equivalent to $y$. After that, signal $x$ is obtained from $s$ by inverse \textit{sparsification}, i.e., $x= \Psi s$.

The approach of reconstructing sparse signal by solving \eqref{eq: CS reconstruction l1} is referred as a convex optimization method. Some other approaches such as LASSO \cite{ADMMSBoyd, Lasso}, OMP \cite{Davis}, CoSaMP \cite{Needell}, SpaRSA \cite{Stephen}, etc. \cite{David} can be used to reconstruct the sparse signal from the measurements. 

\section{Proposed Blind Multi-Image Steganography Scheme}\label{propose method}
Our proposed blind multi-image steganography scheme, which is based on $\ell_1$-minimization and sparse approximation, consists of embedding secret images and their extraction from the generated stego-image. These parts are discussed in the respective subsections below.
\subsection{Secret Images Embedding}\label{secret images embedding}
First, we perform sub-sampling on a cover image and obtain four sub-sampled images (or sub-images). Let $CI$ is the cover image of size $N\times N$, then the four sub-images are obtain as
\begin{equation}
\begin{aligned}\label{eq:sampling}
& {CI}^1(n_1,n_2) = CI(2n_1-1,2n_2-1),   \\ %\nonumber 
& {CI}^2(n_1,n_2) = CI(2n_1,2n_2-1),   \\
& {CI}^3(n_1,n_2) = CI(2n_1-1,2n_2),   \\
& {CI}^4(n_1,n_2) = CI(2n_1,2n_2),   
\end{aligned}
\end{equation}
where $n_1,n_2 = 1,2,\hdots, \frac{N}{2}$ (in our case, $N$ is completely divisible by $2$); $CI^k$, $\text{for} \; k=\{1,2,3,4\}$, are the four sub-images; and $CI(\cdot,\cdot)$ is the pixel value at $(\cdot,\cdot)$.

Originally, these sub-images are not sparse; hence, next, we perform block-wise sparsification of each of these images. For this, we divide each sub-image into blocks of size $b\times b$ and obtain $\frac{N^2}{4\times b^2}$ blocks for each sub-image (in our case, $b$ completely divides $N$). Now, we consider each block as a vector of size $b^2\times 1$, and then sparsify them (as in \eqref{eq:CS sparsification}) using discrete cosine transform matrix of size $b^2\times b^2$ as the sparsification matrix $\Psi$. That is,
\begin{equation}\label{eq:propose sparsification}
s_i = \Psi^T x_i,
\end{equation}  
where $i=1,2,\hdots,\frac{N^2}{4\times b^2}$, $x_i$ and $s_i$ are the $i^{th}$ original and sparse vector representation of the respective blocks, and $\Psi^T$ is the transpose of $\Psi$. Now, we consider these sparse vector in the zig-zag scanning order as given in one of our previous paper.\cite{Steg_Rohit} As a consequence of sparsification, each sparse vector has few coefficients of significant values and the remaining coefficients of very small or zero values. Thus, we categories each vector into two groups $s_{i,u}\in R^{p_1}$ and $s_{i,v}\in R^{p_2}$, where $p_1$ and $p_2$ are the number of coefficients having large values and small values (or zero values), respectively, and $p_1 + p_2 = b^2$. Now, we project each sparse vector onto linear measurements as 
\begin{equation}\label{eq: propose measurements}
y_i=\begin{bmatrix}y_{i,u}\\ y_{i,v} \end{bmatrix}=
\begin{bmatrix}
s_{i,u}\\ \Phi s_{i,v}	
\end{bmatrix},
\end{equation}
where $\Phi$ is the measurement matrices, which is a $m\times p_2$ (for our case, $m>p_2$) matrix of normally distributed random numbers; and $y_i\in R^{({p_1+m})\times 1}$ is the set of linear measurements. Since the distribution of coefficients of the generated sparse vectors is almost the same for all blocks of an image, we use the same measurement matrix for all blocks. 
%Note that we perform the above sparsification and projection onto linear measurements for all blocks of each sub-image. 

Next, we perform processing in the secret images for embedding them into the cover image. In our proposed steganography scheme, we can embed a maximum of four secret images, one in each of the four sub-images of a single cover image. If we want to embed less than four secret images, we randomly select numbers of sub-images equal to the number of secret images that we want to embed, from the total four sub-images. Let $S^k$, $\text{for}  \; k=\{1,2,3,4\}$, are the four secret images each of the size $M\times M$. First, we perform block-wise DCT to each of these images and obtain their corresponding DCT coefficients. Here, the size of each block is $l\times l$, and hence, we have $\frac{M^2}{l^2}$ number of blocks for each secret image (in our case, $l$ completely divides $M$). Now, we consider these DCT coefficients as a vector in the zig-zag scanning order as given in \cite{Steg_Rohit}. Let $t^{k}_{i} \in R^{l^2\times 1}$, for $i=1,2,\hdots,\frac{M^2}{l^2}$, be the vector representation of the DCT coefficients of the $S^k$ secret image.

Now, we perform embedding of the secret images in the cover image. We embed $t_i$ DCT coefficients from the secret image into $y_i$ linear measurements of one sub-image of the cover image. This embedding is given in \textbf{Algorithm \ref{alg:Embedding rule}} that proposes the embedding rule. In this, we show embedding of only one secret image into one sub-image.
%...................Embedding Rule Algorithm.......................%
{
	\begin{algorithm}[h]
	%\scriptsize %\small, \footnotesize, \scriptsize, or \tiny
		\caption{Embedding Rule}\label{alg:Embedding rule}
		\begin{algorithmic}[1]
			\renewcommand{\algorithmicrequire}{\textbf{Input:}}
			\renewcommand{\algorithmicensure}{\textbf{Output:}}
			\Require \quad
			\begin{itemize}
				\item $y$: Sequence of linear measurements of the cover image. 
				\item $t$: Sequence of transform coefficients of the secret image.
				\item The value of $p_1$, $p_3$, $\alpha$, $\beta$, $\gamma$ and $c$ (discuss in Sections \ref{secret images embedding} and \ref{experimental results}).
			\end{itemize} 
			
			\Ensure \quad
			\begin{itemize}
				\item ${y}'$: The modified version of the linear measurements.
			\end{itemize}
		%	\\ \textit{Initialisation}: 
		    \State Initialize ${y'}$ to $y$
			\For {$i = 1$ to $\frac{N^2}{4\times b^2}$}\label{alg:eq 1 embedding rule} 
		          \begin{align}
                     & {y'}_i(p_1) = y_i(p_1-2c) + \alpha \times t_i(1).  \nonumber  
                  \end{align}
                \For {$j = p_1 - c + 1$ to $p_1 - 1$}\label{alg:eq 2 embedding rule} 
                     \begin{align}
                     & {y'}_i(j) = y_i(j-c) + \beta \times t_i(j-p_1+c+1).  \nonumber  
                  \end{align}
                \EndFor  
                \For {$k = p_1 + p_3 + 1$ to $p_1 + 2\times p_3 - c$}\label{alg:eq 3 embedding rule}
                     \begin{align}
                     & {y'}_i(k) = y_i(k - p_3 + c) + \gamma \times t_i(k - p_1 - p_3 + c).  \nonumber  
                  \end{align}
                \EndFor  
			\EndFor		\\		
			\Return ${y}'$
		\end{algorithmic}
	\end{algorithm}
}

In this algorithm, $p_1$ represents the number of coefficients having large values (as discussed above), $p_3$ is the number of DCT coefficients from each $t_i$ that are embedded into the cover sub-image, and $\alpha$, $\beta$, $\gamma$ \& $c$ are the constants. Here, the steps number \ref{alg:eq 1 embedding rule}, \ref{alg:eq 2 embedding rule} and \ref{alg:eq 3 embedding rule} shows the embedding of the first coefficient, the next $c-1$ coefficients, and the remaining coefficients, respectively. Note that here, we choose all the parameters such that $\frac{M^2}{l^2}$ will be less than or equal to $\frac{N^2}{4\times b^2}$, $p_3$ coefficient from $t_i$ can be embedded into $y_i\in R^{p_1+m}\times 1$ using \textbf{Algorithm \ref{alg:Embedding rule}}, and the stego-images \& the extracted secret images preserve good visual quality. We discuss the values of all these parameters in Experimental Results section (i.e., in section \ref{experimental results}).

Finally, we construct the stego-image. As earlier, we can embed a maximum of four secret images into four sub-images of a single cover image. Hence, we first construct four sub-stego-images and then perform inverse sampling to obtain a stego-image from these four sub-stego-images. Let $s'_i$ be the sparse vector of the $i^{th}$ block of a sub-stego-image (say $k^{th}$), then 
\begin{equation}
\begin{aligned}\label{eq:stego reconstruction}
& s'_{i,u} = y'_i(1:p_1) \: \text{and} \\
& s'_{i,v} = \min_{s'_{i,v}} \left \|{s'_{i,v}} \right \|_{1} \\
& \text{Subject to} \hspace{0.1cm} \Phi {s'_{i,v}} = y'_i(p_1+1:p_1+m).
\end{aligned}
\end{equation}
where ${y}'_i(a:b)$ is the range from $a^{th}$ element to $b^{th}$ element of the vector ${y}'_i$. For the solution of the minimization problem of \eqref{eq:stego reconstruction}, we use LASSO (least absolute shrinkage and selection operator) formulation of it and then solve it using ADMM (alternating direction method of multipliers) algorithm \cite{ADMMSBoyd, Lasso}. The reason for this is that it has a wide application in the image processing domain. 
%(see \cite{ADMMSBoyd} and \cite{Lasso} for detail of LASSO and ADMM, respectively). 
The sparse vector $s'_i$ is the concatenation of $s'_{i,u}$ and $s'_{i,v}$. After that, we perform inverse sparsification and obtain non-sparse vectors as $x'_i = \Psi s'_i$. Now, we covert each vector $x'_i$ into block of size $b\times b$, and then construct the sub-stego-image of size $\frac{N}{2} \times \frac{N}{2}$ by arranging all these blocks. In the end, we perform inverse sampling (see \eqref{eq:sampling}) and obtain a single stego-image from the four sub-stego-images.

\subsection{Secret Images Extraction}\label{secret images extraction}
In this subsection, we discuss the process of extraction of the secret images from the stego-image. Initially, we perform sampling (as done in subsection \ref{secret images embedding} using \eqref{eq:sampling}) onto the stego-image to obtain four sub-stego-images. Let $T^k$, for $k=\{1,2,3,4\}$, are the four sub-stego-images. Since the extraction of all the secret images is similar, here, we discuss the extraction of only one secret image from one sub-stego-image. In this process, first, we perform block-wise sparsification of the sub-stego-image. For this, we divide this image into blocks of size $b\times b$ and then consider each block as a vector of $b^2\times 1$. Here, we have a total of $\frac{N^2}{4\times b^2}$ blocks for a sub-stego-image. Next, we sparsified each vector (say $x''_i$), as done in subsection \ref{secret images embedding}, using the same sparsification matrix and \eqref{eq:propose sparsification}, and then obtain sparse vector (say $s''_i$). 

Next, as earlier, we consider these sparse vector in the zig-zag scanning order, and then categories each vector into two groups $s''_{i,u}\in R^{p_1}$ and $s''_{i,v}\in R^{p_2}$, where as earlier, $p_1$ and $p_2$ are the number of coefficients having large values and small values (or zero values), respectively. After that, we project each sparse vector onto linear measurements (say $y''_i\in R^{(p_1+m)\times 1}$), as done in subsection \ref{secret images embedding}, using the same measurement matrix $\Phi\in R^{m\times p_2}$ and \eqref{eq: propose measurements}. This ${y''}$ have DCT coefficients of the secret image that is extracted by the extraction rule given in \textbf{Algorithm \ref{alg:Extraction rule}}. This extraction rule is reverse of the embedding rule, given in \textbf{Algorithm \ref{alg:Embedding rule}}.
%...................Embeextraction Rule Algorithm.......................%
{
	\begin{algorithm}[h]
	%\scriptsize %\small, \footnotesize, \scriptsize, or \tiny
		\caption{Extraction Rule}\label{alg:Extraction rule}
		\begin{algorithmic}[1]
			\renewcommand{\algorithmicrequire}{\textbf{Input:}}
			\renewcommand{\algorithmicensure}{\textbf{Output:}}
			\Require \quad
			\begin{itemize}
				\item ${y''}$: Sequence of linear measurements of the stego-image. 
				\item The value of $p_1$, $p_3$, $\alpha$, $\beta$, $\gamma$ and $c$ (discuss in Sections \ref{secret images embedding} and \ref{experimental results}).
			\end{itemize} 
			
			\Ensure \quad
			\begin{itemize}
				\item $t'$: Sequence of transform coefficients of the extracted secret image.
			\end{itemize}
		%	\\ \textit{Initialisation}: 
		    \State Initialize ${t'}$ to zeros
			\For {$i = 1$ to $\frac{N^2}{4\times b^2}$}
		          \begin{align}          
                     & {t'}_i(1) = \frac{y''_i(p_1) - y''_i(p_1-2c)}{\alpha} .  \nonumber  
                  \end{align}
                \For {$j = p_1 - c + 1$ to $p_1 - 1$}
                     \begin{align}          
                     & {t'}(j-p_1+c+1) = \frac{y''_i(j) - y''_i(j-c)}{\beta}.  \nonumber  
                  \end{align}
                \EndFor  
                \For {$k = p_1 + p_3 + 1$ to $p_1 + 2\times p_3 - c$}
                     \begin{align}          
                     & {t'}(k - p_1 - p_3 + c) = \frac{y''_i(k) - y''_i(k - p_3 + c)}{\gamma}.  \nonumber  
                  \end{align}
                \EndFor  
			\EndFor		\\		
			\Return ${t'}$
		\end{algorithmic}
	\end{algorithm}
}

In \textbf{Algorithm \ref{alg:Extraction rule}}, the size of $t'_i$ is same as the size of $t_i$, i.e., $l^2\times 1$. After extracting DCT coefficients ${t'}$ from the stego-image, we convert each vector $t'_i$ into the blocks of size $l\times l$, and then perform block-wise inverse discrete cosine transformation (IDCT) to obtain secret image pixels. Finally, we obtain extracted secret image of size $M\times M$ by arranging all these blocks. 

As mentioned earlier, this steganography scheme is a blind multi-image steganography scheme because it does not require any cover image data at the receiver side for the extraction of secret images.

\section{Experimental Results}\label{experimental results}
\begin{figure}[bh]
	\centering
	\begin{subfigure}[b]{0.17\textwidth}
		\centering
		\includegraphics[width=2.5cm,height=2.5cm,keepaspectratio]{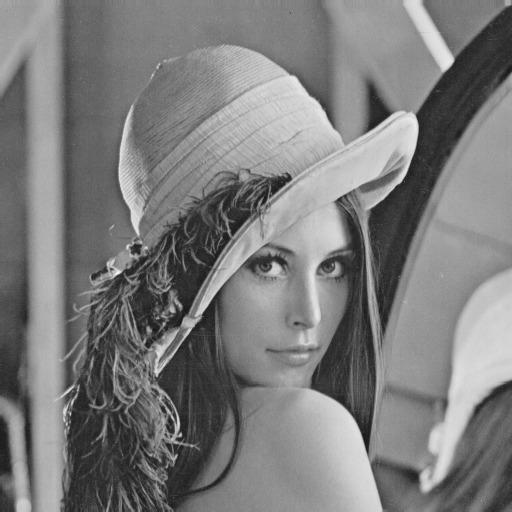}
		\caption{Lena}
		\label{fig:lena}
	\end{subfigure}
	\hfill %
	\begin{subfigure}[b]{0.17\textwidth}
		\centering
		\includegraphics[width=2.5cm,height=2.5cm,keepaspectratio]{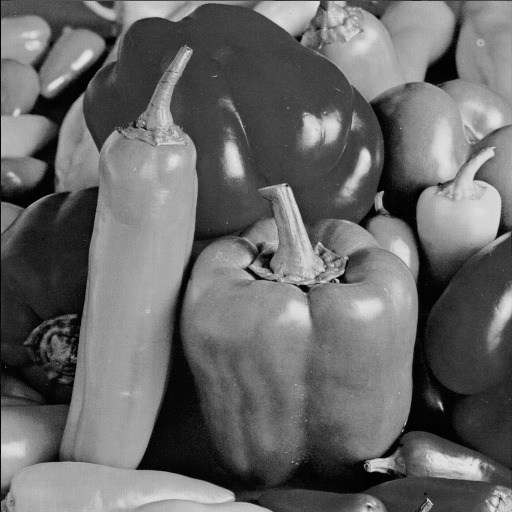}
		\caption{Peppers}
		\label{fig:peppers}
	\end{subfigure}
	\hfill %
	\begin{subfigure}[b]{0.17\textwidth}
		\centering
		\includegraphics[width=2.5cm,height=2.5cm,keepaspectratio]{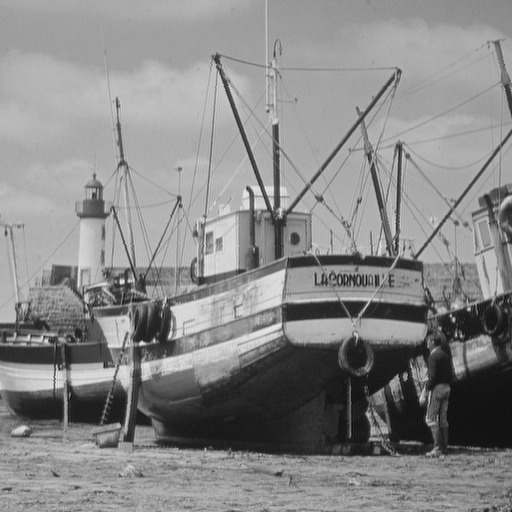}
		\caption{Boat}
		\label{fig:boat}
	\end{subfigure}
	\hfill %
	\begin{subfigure}[b]{0.17\textwidth}
		\centering
		\includegraphics[width=2.5cm,height=2.5cm,keepaspectratio]{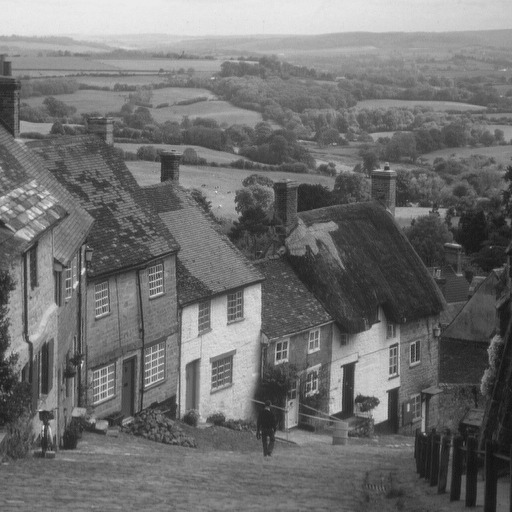}
		\caption{Goldhill}
		\label{fig:goldhill}
	\end{subfigure}
	\hfill %
	\begin{subfigure}[b]{0.17\textwidth}
		\centering
		\includegraphics[width=2.5cm,height=2.5cm,keepaspectratio]{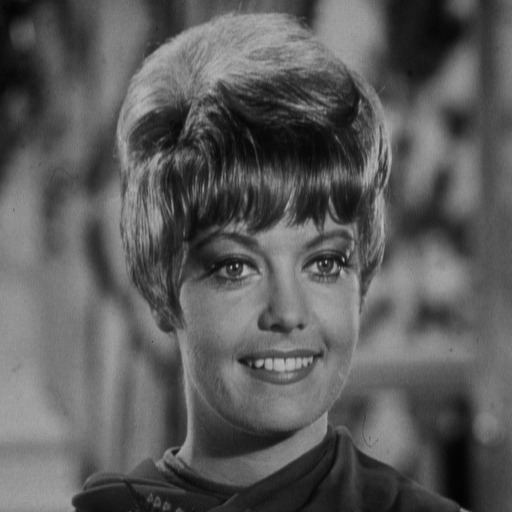}
		\caption{Zelda}
		\label{fig:zelda}
	\end{subfigure}
	\hfill %
	\begin{subfigure}[b]{0.17\textwidth}
		\centering
		\includegraphics[width=2.5cm,height=2.5cm,keepaspectratio]{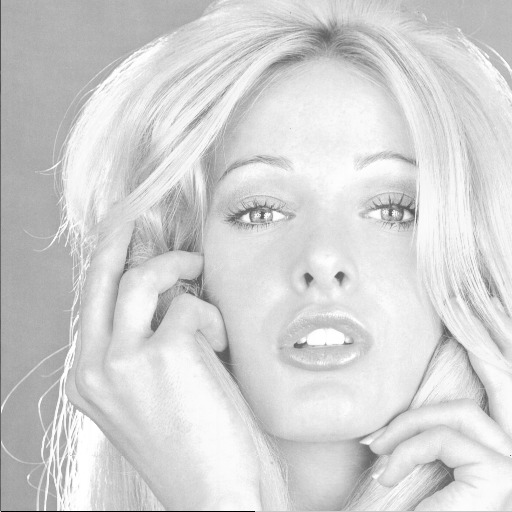}
		\caption{Tiffany}
		\label{fig:tiffany}
	\end{subfigure}
	\hfill %
	\begin{subfigure}[b]{0.17\textwidth}
		\centering
		\includegraphics[width=2.5cm,height=2.5cm,keepaspectratio]{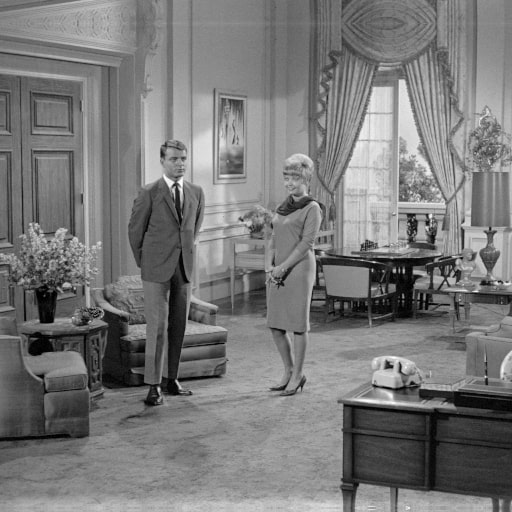}
		\caption{Liv. room}
		\label{fig:livingroom}
	\end{subfigure}
	\hfill %
		\begin{subfigure}[b]{0.17\textwidth}
		\centering
		\includegraphics[width=2.5cm,height=2.5cm,keepaspectratio]{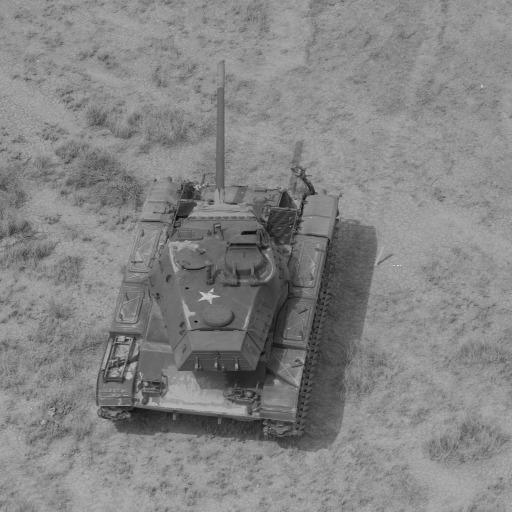}
		\caption{Tank}
		\label{fig:tank}
	\end{subfigure}
	\hfill %
	\begin{subfigure}[b]{0.17\textwidth}
		\centering
		\includegraphics[width=2.5cm,height=2.5cm,keepaspectratio]{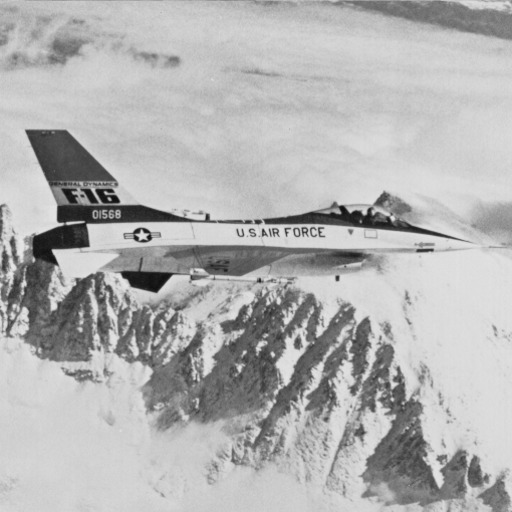}
		\caption{Airplane}
		\label{fig:jetplane}
	\end{subfigure}
	\hfill %
	\begin{subfigure}[b]{0.17\textwidth}
		\centering
		\includegraphics[width=2.5cm,height=2.5cm,keepaspectratio]{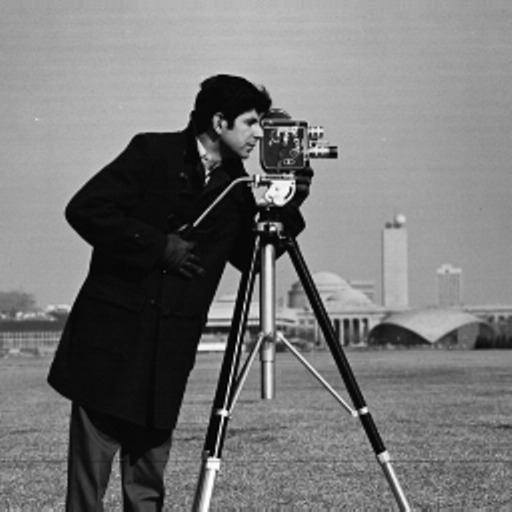}
		\caption{Cameraman}
		\label{fig:cameraman}
	\end{subfigure}
	\caption{Test images used in our experiments}
	\label{Figure:test images}
   \end{figure}

Experiments are carried out in $MATLAB^{\textregistered}$ on a machine with Intel Core i3 processor @2.30 GHz and 4GB RAM. We use a set of standard gray-scale images to test our $\ell_1$SABMIS. Some sample test images used in our experiments are shown in Fig. \ref{Figure:test images}. These images are taken from the USC-SIPI image database \cite{SIPI}, and have varying texture property. In this paper, we take all these ten images as the cover images, and four images (Fig. \ref{fig:lena}, Fig. \ref{fig:peppers}, Fig. \ref{fig:zelda} and Fig. \ref{fig:cameraman}) as the secret images for our experiments. However, we can use any of the ten images apart from these four as the secret image.
%%%%%%%%%%%\begin{figure*}[!b]

Though the images shown in Fig. \ref{Figure:test images} seems to be of same size, however for our experiments, the size of each cover image is kept as $1024\times 1024$ (i.e., $N\times N$), and the size of each secret image is kept as $512\times 512$ (i.e., $M\times M$). We take blocks of size $8\times 8$ for both the cover images and the secret images (i.e., $b\times b$ and $l\times l$). Recall from subsection \ref{secret images embedding}, the size of measurement matrix $\Phi$ is $m\times p_2$, where $p_1+p_2=b^2$ (here, $b^2 = 64$). Usually, for most images, more than half of the coefficient in a DCT sparsified vector has value either very small or zero. Hence, in our experiments, we take $p_1 = p_2 = 32$, and $m = 10\times p_2$ (i.e, we over-sampled linear measurements). In general, the DCT coefficients can be divided into three sets, low frequencies, middle frequencies, and high frequencies. Low frequencies are associated with the illumination, middle frequencies are associated with the structure, and high frequencies are associated with the noise or small variation details. We discarded these high-frequency coefficients, which are usually half of the total coefficients. Hence, we take $p_3 = 32$. Due to the property of DCT, the first coefficient have the highest value, then a few coefficients have values lower than the first one, and remaining coefficients have values lowest. These three category of coefficients are embedded in the step numbers \ref{alg:eq 1 embedding rule}, \ref{alg:eq 2 embedding rule} and \ref{alg:eq 3 embedding rule} of \textbf{Algorithm \ref{alg:Embedding rule}}, respectively. Hence, to obtain good quality extracted secret image, we take $\alpha=0.01$, $\beta=0.1$, $\gamma=1$, and $c=\{6,8\}$ (the results are reported for only one value of $c$ that gave better performance). 

A successful steganography scheme should have high embedding capacity, and should not distort the cover media significantly (i.e., the distortion should be visually imperceptible). Thus, we evaluate the performance of our proposed $\ell_1$SABMIS by analyzing these two metrics. As this paper proposes a multi-image steganography scheme, we evaluate the performance of $\ell_1$SABMIS for the cases where all the four images and less than four images are embedded. 

%\subsection{Embedding Capacity Analysis}
The embedding capacity (or embedding rate) is the number (or length) of secret information bits that can be embedded in each pixel of the cover image. It is measured in bits per pixel (bpp). % and calculated as follows
%\begin{equation}
%\text{EC in bpp} = \frac{\text{Total number of embedded bits}}{\text{Total numbe of pixels in the cover image}}.
%\end{equation}
%This embedding capacity is also measured in percentage, which is calculated as follows 
%\begin{equation}
%\text{EC in \%} = \frac{\text{EC in bpp}}{\text{Size of each pixel in bits}}.
%\end{equation}
%In this scheme, we perform experiments on gray-scale images, and we embedded maximum four secret images of size $512\times 512$ in a cover image of size $1024\times 1024$. 
Thus, we have embedding capacities of $2$ bpp, $4$ bpp, $6$ bpp, and $8$ bpp for embedding one, two, three, and four secret images, respectively. Next, we evaluate the quality of the stego-image.
%\subsection{Imperceptibility Analysis}
Imperceptibility is the measure of the invisibility of the secret image that is hidden in the generated stego-image. There is no universal criterion to determine imperceptibility. However, we evaluate it by visual and numerical (PSNR, MSSIM, NCC, NAE, and Entropy) metrics.

\begin{figure}[h]
	\centering
	\begin{subfigure}[b]{0.19\textwidth}
		\centering
		\includegraphics[width=2.5cm,height=2.5cm,keepaspectratio]{zelda.jpg}
		\caption{cover image}
		\label{fig:zelda cover image}
	\end{subfigure}
	\hfill %
%	\begin{subfigure}[b]{0.20\textwidth}
%		\centering
%		\includegraphics[width=3cm,height=2.5cm]{CoverZeldaHist.jpg}
%		\caption{Cover image histogram}
%		\label{fig:zelda cover image hist}
%	\end{subfigure}
%	\hfill %
	\begin{subfigure}[b]{0.19\textwidth}
		\centering
		\includegraphics[width=2.5cm,height=2.5cm,keepaspectratio]{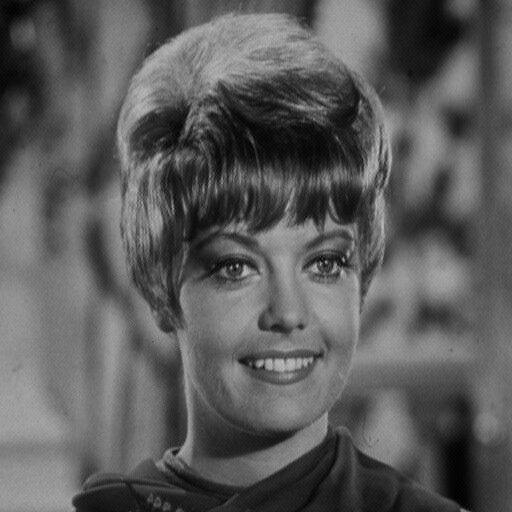}
		\caption{stego-image}
		\label{fig:zelda stego-image}
	\end{subfigure}
	\hfill %
%	\begin{subfigure}[b]{0.20\textwidth}
%		\centering
%		\includegraphics[width=3cm,height=2.5cm]{StegoZeldaHist.jpg}
%		\caption{Stego-image histogram}
%		\label{fig:zelda stego-image hist}
%	\end{subfigure}
%	\hfill
	\begin{subfigure}[b]{0.19\textwidth}
		\centering
		\includegraphics[width=2.5cm,height=2.5cm,keepaspectratio]{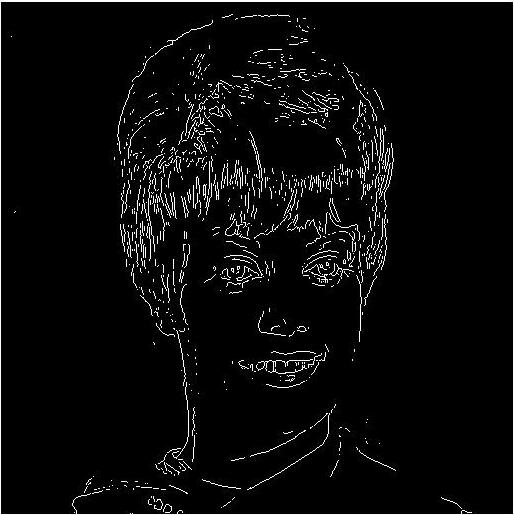}
		\caption{CI edge map}
		\label{fig:zelda cover edge map}
	\end{subfigure}
	\hfill %
	\begin{subfigure}[b]{0.19\textwidth}
		\centering
		\includegraphics[width=2.5cm,height=2.5cm,keepaspectratio]{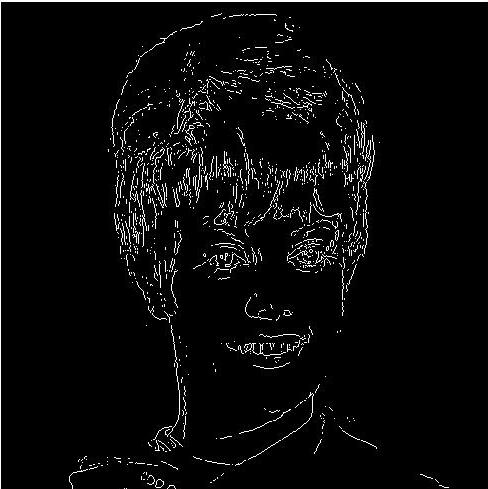}
		\caption{SI edge map}
		\label{fig:zelda stego edge map}
	\end{subfigure}
	%\caption{Edge map of `Pepper' cover image and its stego-image for parameter $|p_1|$=12 and $|m|$=37.}
	%\label{Fig:Edge map}
	\caption{Visual quality analysis between `Zelda' cover image (CI) and its corresponding stego-image (SI)}
	\label{Figure:visual analysis}
\end{figure}

%\subsubsection{Subjective or Visual Measure}\label{Visual similarity}
We construct stego-images corresponding to different test images used in our experiments and then check the distortion visually. We also check their corresponding edge map diagrams. Here, we present the visual comparison only for `Zelda' cover image. This comparison is given in Fig. \ref{Figure:visual analysis}. In this figure, \ref{fig:zelda cover image} shows `Zelda' cover image, \ref{fig:zelda stego-image} shows stego-image, %\ref{fig:zelda cover image hist} shows the histogram of cover image, \ref{fig:zelda stego-image hist} shows the histogram of stego-image, 
\ref{fig:zelda cover edge map} shows the edge map diagram of cover image, and \ref{fig:zelda stego edge map} shows the edge map diagram of stego-image. From these sub-figures, we observe that the stego-image is almost similar to its corresponding cover image. Also, their corresponding edge maps are almost the same. 

\begin{figure}[h]
	\centering
		\includegraphics[width=1\textwidth]{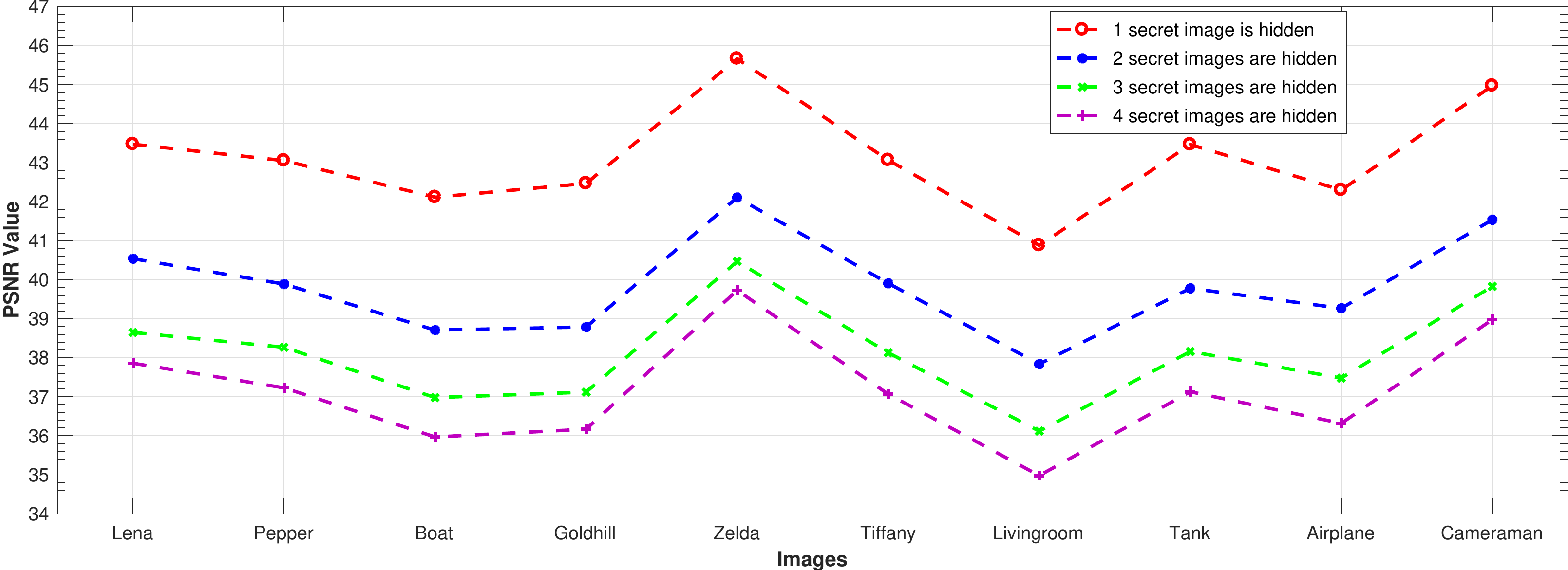}
		\caption{PSNR value of the stego-images when different number of images are hidden}
		\label{fig:PSNRStego1to4Imageshidden}
\end{figure}

\begin{figure}[h]
	\centering
			\includegraphics[width=1\textwidth]{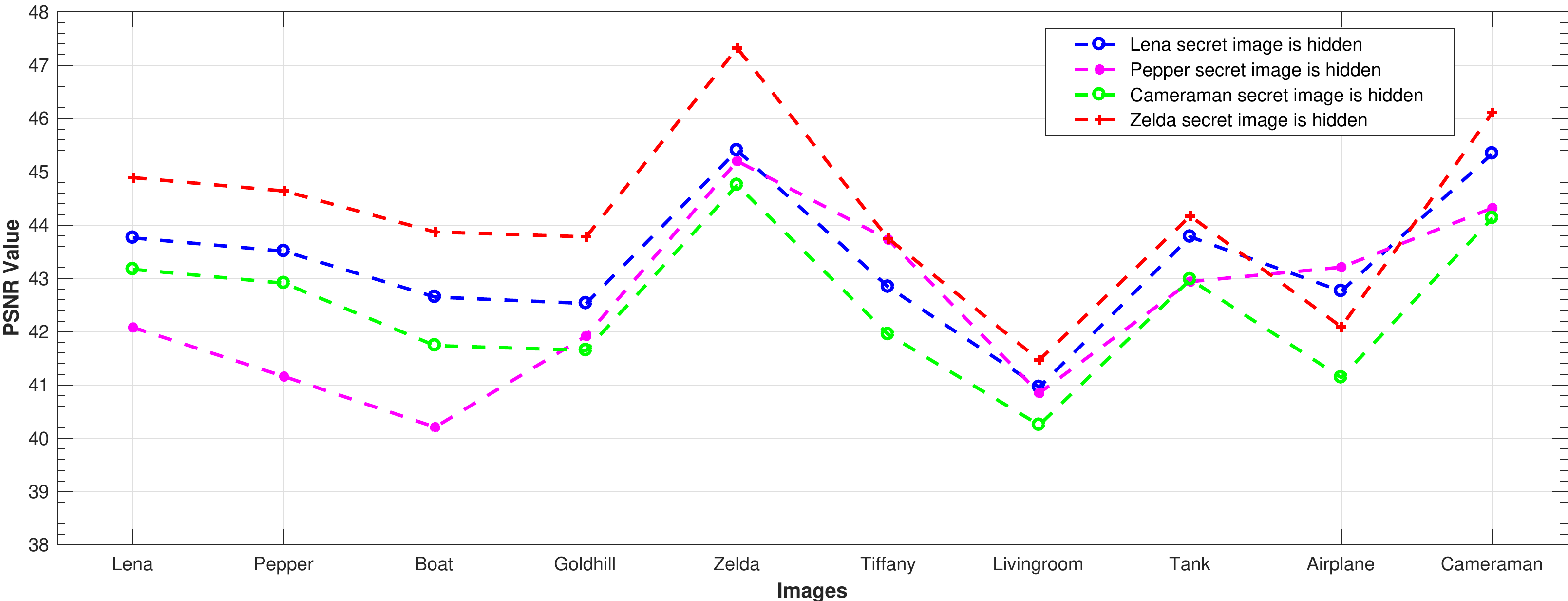}
		\caption{PSNR value of the stego-images when only 1 secret image is hidden}
		\label{fig:PSNRStego1imagehidden}
\end{figure}

%\subsubsection{Objective or Numerical Measures}\label{Numerical similarity}
The numerical metrics evaluate imperceptibility by comparing the cover images and their corresponding stego-images based on some numerical criteria. These include; PSNR value, mean SSIM index, NNC coefficient, NAE, and entropy. 
%\paragraph{\textbf{PSNR:}}\label{PSNR}
PSNR\cite{PSNR_World_Scienticfic} value is evaluated in decibel (dB), and its higher value indicates the higher imperceptibility of the stego-image. In general, value above 30 dB is considered to be good for the quality of the stego-image. \cite{Liu,Zhang2013}  For more details, please see. \cite{Steg_Rohit}
%\begin{equation}%\label{eq:PSNR}
%PSNR=10\log_{10}\frac{{255}^{2}}{MSE}\: dB,
%\end{equation}
%where $MSE$ represents the mean square error between the cover image $CI$ and the stego-image $S'$. The $MSE$ is calculated as:
%\begin{equation}%\label{eq:MSE}
%MSE=\frac{\sum_{i=1}^{r1}\sum_{j=1}^{r2}\left ( CI\left ( i,\, j \right )-S'\left ( i,\, j \right ) \right )^2}{r1\times r2},
%\end{equation}
%where $r1$ and $r2$ represent the row and column numbers of the digital image, respectively. $CI(i,j)$ and  $S'(i,j)$ represent the pixel value of the cover image and the constructed stego-image, respectively. 
The PSNR values of the stego-images corresponding to different test images are given in Fig. \ref{fig:PSNRStego1to4Imageshidden} and Fig. \ref{fig:PSNRStego1imagehidden}. In Fig. \ref{fig:PSNRStego1to4Imageshidden}, we show the PSNR values for the four cases, i.e., embedding one, two, three, and four images. 
%For the case of embedding one, two, or three secret images, the PSNR values shown in this figure are the average value of the PSNR when different combinations of the secret images are embedded.
As we have four secret images, we have the choice of embedding any one of them and present its corresponding PSNR value. However, we embed all four images individually, obtain their PSNR values, and then present the average of these four PSNR values. Similarly, the average PSNR values are presented for the cases when we embed two and three images. For the case of embedding four images, there is only one choice, and hence its corresponding PSNR value is presented. 

In Fig. \ref{fig:PSNRStego1imagehidden}, we show the PSNR values of all the stego-images when all the four secret images are embedded separately. In this figure, we observe the highest PSNR value (i.e., 47.32 dB) when `Zelda' secret image is hidden in `Zelda' cover image, while the lowest PSNR value (i.e., 40.21 dB) when `Peppers' secret image is hidden in `Boat' cover image. Also, we observe that for all the cases, we obtain PSNR values higher than 30 dB, and hence, considered good.

%\paragraph{\textbf{Means SSIM Index:}}\label{SSIM}
We measure the structural similarity between the cover images and their corresponding stego-images by the metric of mean SSIM \cite{SSIM}. The value of mean SSIM index lies between $0$ and $1$, where the value $0$ implies that there is no similarity between the cover image and stego-image, and the value $1$ implies that the cover image is exactly similar to its corresponding stego-image. We measure the degree of similarity between the cover images and their corresponding stego-images by another metric called normalized cross-correlation (NCC) coefficients \cite{NC}. Similar to mean SSIM, the value of NCC equal to $1$ implies that the cover image is exactly similar to its corresponding stego-image. We also measure the quality of stego-image by evaluating NAE \cite{NC} between the cover images and their corresponding stego-images. A value close to $0$ indicates that the stego-image is almost similar to their corresponding cover image. Moreover, we also measure the entropy of the cover images and their corresponding stego-images. Entropy is a statistical randomness measure, which can be used to characterize the texture of an image \cite{Gonzalez}. In Table \ref{Table:Stego-Images Quality}, we give the values of all these metrics for our $\ell_1$SABMIS when hiding all the four secret images. We do not present the values for the cases of embedding less than four secret images as their results will be better than those given in Table \ref{Table:Stego-Images Quality}.

From this table, we observe that all the values of mean SSIM index are equal to $1$. Hence, the cover images and their corresponding stego-images are similar in structure. The values of NCC coefficients are also close to 1. The values of NAE are close to $0$. The values of entropy of stego-images are almost similar to their corresponding cover images. Thus, we can say that the stego-images and their corresponding cover images are almost identical.
%%%%%%%%%%%%%%%%%%Stego-Images Quality Table%%%%%%%%%%%%%%%%%%
\begin{table}[ht]
%\centering
%\caption{Mean SSIM (MSSIM) index, NCC coefficient, NAE, and entropy of the stego-image}\label{Table:Stego-Images Quality}
\tbl{Mean SSIM (MSSIM) index, NCC coefficient, NAE, and entropy of the stego-image.\label{Table:Stego-Images Quality}}
{\begin{tabular}{@{}cccccc@{}} %cccccc
\toprule % \hline
\multicolumn{1}{c}{\multirow{2}{*}{\textbf{\begin{tabular}[c]{@{}c@{}}Cover\\ Image\end{tabular}}}} & \multicolumn{1}{c}{\multirow{2}{*}{\textbf{MSSIM}}} & \multicolumn{1}{c}{\multirow{2}{*}{\textbf{NCC}}} & \multicolumn{1}{c}{\multirow{2}{*}{\textbf{NAE}}} & \multicolumn{2}{c}{\textbf{Entropy}}                                                                                                                                         \\ %\cline{5-6} 
\multicolumn{1}{c}{}                                                                                & \multicolumn{1}{c}{}                                & \multicolumn{1}{c}{}                              & \multicolumn{1}{c}{}                              & \multicolumn{1}{c}{\textbf{\begin{tabular}[c]{@{}c@{}}Original\\ image\end{tabular}}} & \multicolumn{1}{c}{\textbf{\begin{tabular}[c]{@{}c@{}}Stego-\\ image\end{tabular}}} \\ \colrule%\hline
\multicolumn{1}{c}{Lena}                                                                            & \multicolumn{1}{c}{1}                               & \multicolumn{1}{c}{0.9992}                         & \multicolumn{1}{c}{0.009}                         & \multicolumn{1}{c}{7.443}                                                             & \multicolumn{1}{c}{7.463}                                                           \\ %\hline
\multicolumn{1}{c}{Pepper}                                                                          & \multicolumn{1}{c}{1}                               & \multicolumn{1}{c}{0.9997}                        & \multicolumn{1}{c}{0.012}                         & \multicolumn{1}{c}{7.573}                                                             & \multicolumn{1}{c}{7.598}                                                           \\ %\hline
\multicolumn{1}{c}{Boat}                                                                            & \multicolumn{1}{c}{1}                               & \multicolumn{1}{c}{0.9998}                        & \multicolumn{1}{c}{0.012}                         & \multicolumn{1}{c}{7.121}                                                             & \multicolumn{1}{c}{7.146}                                                           \\ %\hline
\multicolumn{1}{c}{Goldhill}                                                                        & \multicolumn{1}{c}{1}                               & \multicolumn{1}{c}{0.9998}                        & \multicolumn{1}{c}{0.013}                         & \multicolumn{1}{c}{7.471}                                                              & \multicolumn{1}{c}{7.486}                                                           \\ %\hline
\multicolumn{1}{c}{Zelda}                                                                           & \multicolumn{1}{c}{1}                               & \multicolumn{1}{c}{0.9964}                        & \multicolumn{1}{c}{0.011}                         & \multicolumn{1}{c}{7.263}                                                             & \multicolumn{1}{c}{7.272}                                                            \\ %\hline
\multicolumn{1}{c}{Tiffany}                                                                         & \multicolumn{1}{c}{1}                               & \multicolumn{1}{c}{0.9999}                        & \multicolumn{1}{c}{0.006}                         & \multicolumn{1}{c}{6.602}                                                             & \multicolumn{1}{c}{6.628}                                                           \\ %\hline
\multicolumn{1}{c}{Livingroom}                                                                      & \multicolumn{1}{c}{1}                               & \multicolumn{1}{c}{0.9996}                        & \multicolumn{1}{c}{0.013}                         & \multicolumn{1}{c}{7.431}                                                             & \multicolumn{1}{c}{7.438}                                                           \\ %\hline
\multicolumn{1}{c}{Tank}                                                                            & \multicolumn{1}{c}{1}                               & \multicolumn{1}{c}{0.9998}                        & \multicolumn{1}{c}{0.014}                         & \multicolumn{1}{c}{6.372}                                                             & \multicolumn{1}{c}{6.405}                                                           \\ %\hline
\multicolumn{1}{c}{Airplane}                                                                        & \multicolumn{1}{c}{1}                               & \multicolumn{1}{c}{0.9972}                        & \multicolumn{1}{c}{0.015}                         & \multicolumn{1}{c}{6.714}                                                              & \multicolumn{1}{c}{6.786}                                                           \\ %\hline
\multicolumn{1}{c}{Cameraman}                                                                       & \multicolumn{1}{c}{1}                               & \multicolumn{1}{c}{1.0000}                             & \multicolumn{1}{c}{0.009}                         & \multicolumn{1}{c}{7.055}                                                             & \multicolumn{1}{c}{7.123}                                                           \\ \midrule%\hline
\multicolumn{1}{c}{\textbf{Average}}                                                                & \multicolumn{1}{c}{\textbf{1}}                      & \multicolumn{1}{c}{\textbf{0.9991}}              & \multicolumn{1}{c}{\textbf{0.011}}                & \multicolumn{1}{c}{\textbf{7.104}}                                                    & \multicolumn{1}{c}{\textbf{7.1343}}                                                 \\ \botrule%\hline
\end{tabular}}
\end{table}
%%%%%%%% Security Analysis %%%%%%%%%%%%%
%\subsection{Security Analysis}\label{Security Analysis}
In addition to high embedding capacity with good quality stego-image, $\ell_1$SABMIS is also resistant against \textit{steganographic} attacks. The reason for this is that $\ell_1$SABMIS is a transformed domain-based technique. And, in this scheme, we embed secret images by an indirect embedding strategy {(i.e., LSB flipping based embedding is not adopted)}. 
%Hence, it makes no sense to consider \textit{steganographic} attacks \cite{Westfeld, PM1_steganography}.
%%%%%%%%%%%%%%%%% Comparison with existing methods %%%%%%%
\subsection{Performance Comparison}\label{Performance Comparison}
In this subsection, we compare the performance of our $\ell_1$SABMIS with the existing steganography schemes. These results are given in Table \ref{Table:Performance comparison with various other steganography schemes}. 
%%%%%%%%%%%%%%%%%%%%%%%%%%%%%%%%%%%%Table Performance comparson%%%%%%%%%%%%%%%%%%%%%%
\begin{table}[ht]
%\centering
%\caption{Performance comparison of our $\ell_1$SABMIS with various other steganography schemes.}
%\label{Table:Performance comparison with various other steganography schemes}
%\setlength{\tabcolsep}{4pt}
\tbl{Performance comparison of our $\ell_1$SABMIS with various other steganography schemes.\label{Table:Performance comparison with various other steganography schemes}}
{\begin{tabular}{@{}ccccc@{}} % |c|c|c|c|c|
\toprule % \hline
\multicolumn{2}{c}{\textbf{\begin{tabular}[c]{@{}c@{}}Steganography\\ Scheme\end{tabular}}} & \textbf{\begin{tabular}[c]{@{}c@{}}Embedding\\ Capacity\\ (in bpp)\end{tabular}} & \textbf{\begin{tabular}[c]{@{}c@{}}PSNR\\ (in dB)\end{tabular}} & \textbf{\begin{tabular}[c]{@{}c@{}}Resistant to \\ Steganographic \\ Attacks?\end{tabular}} \\ \colrule % \hline
\multicolumn{2}{c}{\cite{S_Hemalatha}}                                                              & 1.33                                                                             & 44.75                                                           & Yes                                                                                         \\ %\hline
\multicolumn{2}{c}{\cite{Prashanti}}                                                              & 2                                                                                & 46.36                                                           & No                                                                                          \\ %\hline
\multicolumn{2}{c}{\cite{Sanjutha_MK}}                                                              & 2.67                                                                             & 48.25                                                           & Yes                                                                                   \\ %\hline
\multicolumn{2}{c}{\cite{S_Arunkumar_MedicalImage_JIFS}}                                                               & 0.5                                                                              & 51.15                                                           & Yes                                                                                         \\ %\hline
\multicolumn{2}{c}{\cite{S_Arunkumar_MedicalImage}}                                                               & 0.25                                                                             & 49.69                                                           & Yes                                                                                         \\ %\hline
\multirow{4}{*}{\textbf{$\ell_1$SABMIS}}                            & \textbf{Hide 1 Image}                            & \textbf{2} & \textbf{44.98} & \multirow{4}{*}{\textbf{Yes}}                                                                        \\ %\cline{2-4}
 & \textbf{Hide 2 Images} & \textbf{4}  & \textbf{42.54} &                                                                     \\ %\cline{2-4}
& \textbf{Hide 3 Images} & \textbf{6} & \textbf{39.83}          &                                                                                             \\ %\cline{2-4}
& \textbf{Hide 4 Images} & \textbf{8} & \textbf{38.98}                                                           &                                                                                             \\ \botrule %\hline
\end{tabular}}
%\label{Table:Performance comparison with various other steganography schemes}
\end{table}

In this table, the first column represents various steganography schemes, and the remaining columns represent the metrics used for the comparison. As earlier, a successful steganography scheme should have high embedding capacity with a considerably good quality stego-image. Also, it should be resistant to steganographic attacks. Hence, we use embedding capacity, PSNR value, and checking which schemes are resistant to steganographic attacks or not as the performance metrics for the comparison. In this table, the embedding capacity is represented in bit per pixel, where each pixel is considered as a gray-scale. For these existing schemes, the data is not available for all the test images used in this experiment. Hence, in this table,  we report the average PSNR values of all the images given in the respective papers.

From this table, we observe that except \cite{Prashanti} and \cite{Sanjutha_MK}, our $\ell_1$SABMIS outperforms all other existing steganography schemes. The scheme proposed in \cite{Prashanti} is based on LSB based embedding, which is not resistant to steganographic attacks. The scheme proposed in \cite{Sanjutha_MK} embeds single secret image in a color image, which is different from our goal. Also, it has limited embedding capacity. Hence, we can say that out of all these schemes, $\ell_1$SABMIS embeds multiple secret images with high embedding capacity and good quality stego-image. Also, $\ell_1$SABMIS is resistant to steganographic attacks. 

As discussed earlier, we also calculate the mean SSIM index, NCC coefficients, NAE, and entropy values for the performance evaluation. However, these results are not compared with other techniques because the data for the same are not available for these other techniques. 
%%%%%%%%% Secret Recover Quality Assesment %%%%%%%%%
\subsection{Quality Assesment of Secret Recovered Image}\label{Quality Assesment of Secret Recovered Image}
%As earlier, in $\ell_1$SABMIS, four secret images can be hidden in a cover image and obtained stego-image is shared with the receiver. Hence, for a good steganography scheme, these extracted secret images must not be distorted too much. 
As earlier, PSNR value can be used to assess the quality of the secret recovered image, and the value greater than 30 dB is considered good. However, this is not true for every case. For example, the steganography scheme proposed in \cite{S_Hemalatha} has PSNR values greater than 30 dB for the extracted secret images. However, these images have black colored and white colored alternate horizontal and vertical lines. Hence, we only report other metrics to evaluate the quality of the extracted/ recovered secret image.

\begin{figure}[h]
	\centering
	\begin{subfigure}[b]{0.19\textwidth}
		\centering
		\includegraphics[width=2.5cm,height=2.5cm,keepaspectratio]{Peppers.jpg}
		\caption{`Peppers' secret image}
		\label{fig:peppers secret image}
	\end{subfigure}
	\hfill %
	\begin{subfigure}[b]{0.19\textwidth}
		\centering
		\includegraphics[width=2.5cm,height=2.5cm,keepaspectratio]{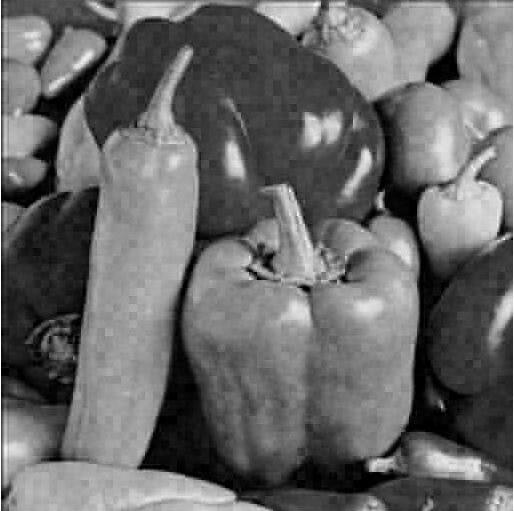}
		\caption{Extracted image}
		\label{fig:peppers secret recovered image}
	\end{subfigure}
	\hfill %
	\begin{subfigure}[b]{0.19\textwidth}
		\centering
		\includegraphics[width=2.5cm,height=2.5cm,keepaspectratio]{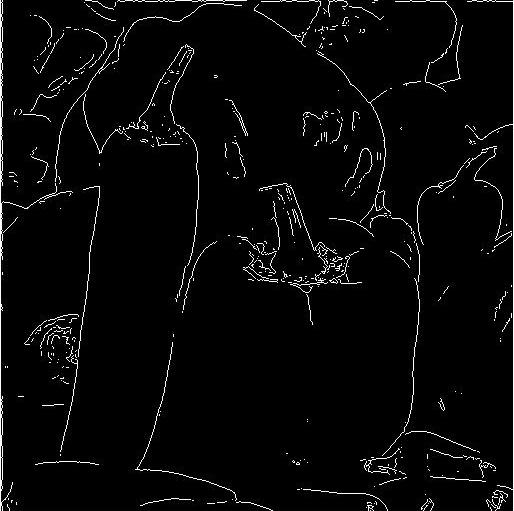}
		\caption{Secret image edge map}
		\label{fig:peppers secret edge map}
	\end{subfigure}
	\hfill %
	\begin{subfigure}[b]{0.19\textwidth}
		\centering
		\includegraphics[width=2.5cm,height=2.5cm,keepaspectratio]{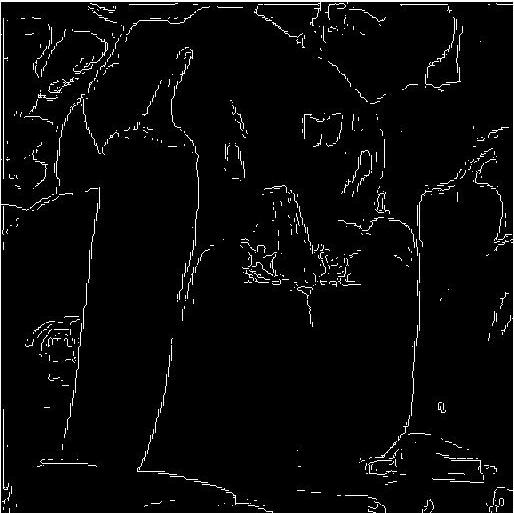}
		\caption{Extracted image edge map}
		\label{fig:peppers secret recovered edge map}
	\end{subfigure}
	\caption{Visual quality analysis between 'Peppers' secret image and `Peppers' extracted image from `Zelda' stego-image.}
		\label{Figure:secret recovered image visual analysis}
\end{figure}

As human is the final spectator of the extracted secret image, human observers are considered the final arbiter to assess the quality of these images. In Fig. \ref{fig:peppers secret image} and Fig. \ref{fig:peppers secret recovered image}, we show the `Pepper' secret image and the extracted secret image from `Zelda' stego-image. From these figures, we observe that there is very little distortion in the extracted image. Besides this, we also show their corresponding edge maps diagram in Fig. \ref{fig:peppers secret edge map} and \ref{fig:peppers secret recovered edge map}, respectively. Again, we observe very little variation in their corresponding edge maps diagrams. 
%%%%%%%%%%%%%%% Figures %%%%%%%%%%%%%

Moreover, we also evaluate the mean SSIM index, NCC coefficient, NAE, and entropy to measure the quality of the extracted secret images. The values of these metrics are given in Table \ref{Table:Secret Image Quality}. From this table, we observe that for all the images, the values of mean SSIM index is $1$, NCC coefficient is close to $1$, NAE is close to $0$, and the entropy of the original secret images and their corresponding extracted secret images are almost the same. Thus, we see that the extracted secret image preserve good quality. 
%%%%%%%%%%%%%%%%%%Table%%%%%%%%%%%%%%%%%%
\begin{table}[ht]
%\centering
%\caption{Mean SSIM (MSSIM) index, NCC coefficient, NAE, and entropy of the extracted secret image}\label{Table:Secret Image Quality}
\tbl{Mean SSIM (MSSIM) index, NCC coefficient, NAE, and entropy of the extracted/ recovered secret image.\label{Table:Secret Image Quality}}
{\begin{tabular}{@{}cccccc@{}}\toprule  %@{}cccccc@{}llllll
%\hline
\multicolumn{1}{c}{\multirow{2}{*}{\textbf{\begin{tabular}[c]{@{}c@{}}Secret\\ Image\end{tabular}}}} & \multicolumn{1}{c}{\multirow{2}{*}{\textbf{MSSIM}}} & \multicolumn{1}{c}{\multirow{2}{*}{\textbf{NCC}}} & \multicolumn{1}{c}{\multirow{2}{*}{\textbf{NAE}}} & \multicolumn{2}{c}{\textbf{Entropy}}                                                                                                                                            \\ %\cline{5-6} 
\multicolumn{1}{c}{}                                                                                 & \multicolumn{1}{c}{}                                & \multicolumn{1}{c}{}                              & \multicolumn{1}{c}{}                              & \multicolumn{1}{c}{\textbf{\begin{tabular}[c]{@{}c@{}}Original\\ Image\end{tabular}}} & \multicolumn{1}{c}{\textbf{\begin{tabular}[c]{@{}c@{}}Recovered\\ Image\end{tabular}}} \\ \colrule%\hline
\multicolumn{1}{c}{Lena}                                                                             & \multicolumn{1}{c}{1}                               & \multicolumn{1}{c}{0.9973}                         & \multicolumn{1}{c}{0.026}                         & \multicolumn{1}{c}{7.446}                                                             & \multicolumn{1}{c}{7.627}                                                              \\ %\hline
\multicolumn{1}{c}{Pepper}                                                                           & \multicolumn{1}{c}{1}                               & \multicolumn{1}{c}{0.9946}                         & \multicolumn{1}{c}{0.329}                         & \multicolumn{1}{c}{7.571}                                                             & \multicolumn{1}{c}{7.624}                                                              \\ %\hline
\multicolumn{1}{c}{Cameraman}                                                                        & \multicolumn{1}{c}{1}                               & \multicolumn{1}{c}{0.9953}                         & \multicolumn{1}{c}{0.027}                         & \multicolumn{1}{c}{7.048}                                                             & \multicolumn{1}{c}{7.258}                                                              \\ %\hline
\multicolumn{1}{c}{Zelda}                                                                            & \multicolumn{1}{c}{1}                               & \multicolumn{1}{c}{0.9972}                         & \multicolumn{1}{c}{0.028}                         & \multicolumn{1}{c}{7.267}                                                             & \multicolumn{1}{c}{7.284}                                                              \\ \midrule%\hline                                                                                     
\multicolumn{1}{c}{\textbf{Average}}                                                                            & \multicolumn{1}{c}{\textbf{1}}                               & \multicolumn{1}{c}{\textbf{0.9961}}                         & \multicolumn{1}{c}{\textbf{0.103}}                         & \multicolumn{1}{c}{\textbf{7.333}}                                                             & \multicolumn{1}{c}{\textbf{7.449}}                                                              \\ \botrule % \hline                                                                                     
\end{tabular}}
%\label{Table:Secret Image Quality}
\end{table}

\section{Conclusions and Future Work}\label{conclusion}
We present a blind multi-image steganography scheme based on $\ell_1$-minimization and sparse approximation. Here, we can embed a maximum of four secret images in a single cover image. Initially, we perform sampling in the cover image and obtain four sub-images. Next, we sparsify each sub-image and then obtain its linear measurements using a random measurement matrix. Finally, using our embedding rule, we embed DCT coefficients of the secret images into the linear measurements. The stego-image is obtained from the modified measurements by solving a $\ell_1$-minimization problem.

We perform experiments on several standard gray-scale images that vary in texture. For performance evaluation, we calculate embedding capacity, PSNR value, mean SSIM index, NCC coefficient, NAE, and entropy. Experiments show that our $\ell_1$SABMIS hides four secret images in a cover image without distorting it significantly and has PSNR values greater than 35 dB, which is usually considered good. Also, the mean SSIM index is equal to $1$, the NCC coefficient is close to $1$, and the value of NAE is close to $0$, which shows that the stego-images are almost identical to its corresponding cover images. We obtain approximately the same entropy value for both the cover images and their corresponding stego-images. We also obtain almost same values for all the metrics (i.e., mean SSIM, NCC coefficients, NAE, and entropy) as above for the extracted secret images. This indicates that the extracted secret images preserve good visual quality. Finally, $\ell_1$SABMIS is also resistant to steganographic attacks. 

In the future, we plan to embed the secret images in other media such as audio, video, etc. We also plan to apply optimization techniques to calculate the values of parameters $\alpha, \beta, \gamma$, etc. used in our embedding and extraction algorithms.

\bibliographystyle{ws-ijwmip}
 
\bibliography{mybib}

\end{document}